\documentclass[a4paper,11pt]{article}
\usepackage{pos}
\usepackage{xspace}
\usepackage{aas_macros}
\usepackage{multirow}
\usepackage{listings}
\usepackage{xcolor}
\usepackage{setspace}

\newcommand{\gammaALPs}{\texttt{gammaALPs}\xspace}
\newcommand{\numpy}{\texttt{numpy}\xspace}
\newcommand{\scipy}{\texttt{scipy}\xspace}
\newcommand{\python}{\texttt{python}\xspace}
\newcommand{\numba}{\texttt{numba}\xspace}
\newcommand{\astropy}{\texttt{astropy}\xspace}

\definecolor{codegreen}{rgb}{0,0.6,0}
\definecolor{codegray}{rgb}{0.5,0.5,0.5}
\definecolor{codepurple}{rgb}{0.58,0,0.82}
\definecolor{backcolour}{rgb}{0.95,0.95,0.92}

\lstdefinestyle{mystyle}{
    backgroundcolor=\color{backcolour},   
    commentstyle=\color{codegreen},
    keywordstyle=\color{magenta},
    numberstyle=\tiny\color{codegray},
    stringstyle=\color{codepurple},
    basicstyle=\ttfamily\footnotesize,
    breakatwhitespace=false,         
    breaklines=true,                 
    captionpos=b,                    
    keepspaces=true,                 
    numbers=left,                    
    numbersep=5pt,                  
    showspaces=false,                
    showstringspaces=false,
    showtabs=false,                  
    tabsize=2
}

\title{\gammaALPs: An open-source python package for computing photon-axion-like-particle oscillations in astrophysical environments}
 \ShortTitle{gammaALPs}

\author*[a]{Manuel Meyer}
\author[b]{James Davies}
\author[a]{Julian Kuhlmann}

\affiliation[a]{Institute for Experimental Physics, University of Hamburg,\\
  Luruper Chaussee 149, 22761 Hamburg, Germany}

\affiliation[b]{University of Oxford, Department of Physics OX1 3RH, Oxford, United Kingdom}

\emailAdd{manuel.meyer@desy.de}

\abstract{
Axions and axion-like particles (ALPs) are hypothetical particles that occur in extensions of the Standard Model and are candidates for cold dark matter. They could be detected through their oscillations into photons in the presence of external electromagnetic fields. \gammaALPs is an open-source python framework that computes the oscillation probability between photons and axions/ALPs. In addition to solving the photon-ALP equations of motion, \gammaALPs includes models for magnetic fields in different astrophysical environments such as jets of active galactic nuclei, intra-cluster and intergalactic media, and the Milky Way. Users are also able to easily incorporate their own custom magnetic-field models. We review the basic functionality and features of \gammaALPs, which is heavily based on other open-source scientific packages such as \numpy and \scipy. Although focused on gamma-ray energies, \gammaALPs can be easily extended to arbitrary photon energies.
}

\FullConference{37$^{\rm{th}}$ International Cosmic Ray Conference (ICRC 2021)\\
		July 12th -- 23rd, 2021\\
		Online -- Berlin, Germany}

\begin{document}
\maketitle

\section{Introduction}
Several extension of the standard model predict the existence of axions and, more generally, axion-like particles (ALPs)~\cite[see, e.g.,][for a review]{2018PrPNP.102...89I}.
Such particles are candidates to explain the cold dark matter content in the Universe and the axion can resolve the strong CP problem of the strong interactions. 
Strategies to search for these particles often utilize their predicted coupling to photons in the presence of electromagnetic fields. The coupling is described through the Lagrangian~\cite{1988PhRvD..37.1237R},
\begin{equation}
\mathcal{L}_{a\gamma} = -\frac 1 4 g_{a\gamma} F_{\mu\nu}\tilde F^{\mu\nu}a
= g_{a\gamma}\mathbf{E}\cdot\mathbf{B}a,
\label{eqn:lagr-alps}
\end{equation}
where $F_{\mu\nu}$ is the electromagnetic field tensor for electric and magnetic fields $\mathbf{E}$ and $\mathbf{B}$, respectively, $\tilde{F}_{\mu\nu}$ is its dual, $a$ is the ALP field strength, and $g_{a\gamma}$ is the photon-ALP coupling constant. 
Hints for the existence of ALPs from the observations of active galactic nuclei (AGNs) at $\gamma$-ray energies spurred the interest in these particles in the astroparticle physics community, however, the evidence remains debated in the literature~\cite[see, e.g.,][for a review]{2016arXiv161107784M}.
Additionally, the absence of oscillation features in $\gamma$-ray (and X-ray) spectra has been used to constrain the coupling strength between ALPs and photons~\cite[see, e.g.,][]{2013PhRvD..88j2003A,2016PhRvL.116p1101A,2020ApJ...890...59R}.

Several authors have studied the oscillations between high-energy photons and ALPs in different astrophysical magnetic field environments~\cite[e.g.,][]{2003JCAP...05..005C,2011PhRvD..84j5030D,2012PhRvD..86g5024H,2014JCAP...09..003M,2008PhRvD..77f3001S,2007PhRvD..76l3011H,2015PhLB..744..375T,2021PhRvD.103b3008D}.
Here, we present the \gammaALPs package, an open-source code written in \python which calculates numerically the photon-ALP oscillation probability in various astrophysical environments. It is based on \numpy~\cite{numpy}, \scipy~\cite{scipy}, and \astropy~\cite{astropy:2013} and uses \numba~\cite{numba} to speed up calculations. 
In addition to various models for magnetic fields along the line of sight to astrophysical sources, it is also possible to provide user-defined models. 
Although originally intended for the application in high-energy $\gamma$-ray astronomy, it can in principle be used for arbitrary photon energies. 

The source code of \gammaALPs is hosted on GitHub\footnote{\url{https://github.com/me-manu/gammaALPs/}} and is licensed under the 3-clause BSD license.
Users can contribute to the code via pull requests. Bug reports and proposals for new functionality can be made through the GitHub issue tracking system. 
The full online documentation is available at \url{https://gammaalps.readthedocs.io}.

\section{Installation}
Up-to-date installation instructions can be found in the online documentation. In a nutshell, the latest release can be installed using the \href{https://pip.pypa.io/en/latest/}{pip} package management tool:  
\begin{lstlisting}[language=Bash]
pip install gammaalps
\end{lstlisting}
Support for an installation through the \href{https://conda.io/docs/index.html}{conda} package manager is foreseen in the near future. A digital object identifier for the \gammaALPs is provided through \href{https://zenodo.org/}{zenodo.org}.

\section{Solving the photon-ALP equations of motion}

The \gammaALPs code solves the equations of motion for photon-ALP oscillations using transfer matrices. 
A detailed solution is provided, e.g., in Refs.~\cite{2010JCAP...05..010B,2011PhRvD..84j5030D} and we provide a summary below (following \cite{2014JCAP...09..003M}) where we highlight important differences and newly implemented features. 
The equations of motion can be derived from the effective Lagrangian, 
\begin{equation}
\mathcal{L} = \mathcal{L}_{a\gamma} + \mathcal{L}_\mathrm{EH} + \mathcal{L}_a,
\end{equation}
where the second term is the effective Euler-Heisenberg Lagrangian and the ALP mass and kinetic terms are
$\mathcal{L}_a = \frac 1 2  \partial_\mu a \partial^\mu a - \frac 1 2 m^2_a a^2$.
Without the loss of generality, in \gammaALPs the propagation direction is assumed to be along the $x_3$ direction. 
ALPs only couple to photons in the presence of a transversal magnetic field $\mathbf{B}_\perp$~\cite{1988PhRvD..37.1237R}, i.e. the $B$-field component in the $\hat{x}_1$-$\hat{x}_2$ plane. 
We let $\psi$ denote the angle that $\mathbf{B}_\perp$ forms with the ${x}_2$ axis. 
Assuming for simplicity $\psi=0$, a monochromatic photon-ALP beam with energy $E$ and photon polarisation states $A_1$ and $A_2$ propagating in a cold plasma filled with a homogeneous magnetic field obeys the equation
 \begin{equation}
\left( i\frac{\mathrm{d}}{\mathrm{d}x_3} + E + \mathcal{M}_0 \right)\Psi(x_3) = 0,
\label{eqn:eom}
\end{equation}
with $\Psi(x_3) = (A_1(x_3),A_2(x_3),a(x_3))^T$ and the mixing matrix $\mathcal{M}_0$,
\begin{equation}
\mathcal{M}_0 = 
\begin{pmatrix}
\Delta_{\perp} & 0 & 0\\
0 & \Delta_{||} & \Delta_{a\gamma} \\
0 & \Delta_{a\gamma} & \Delta_a
\end{pmatrix},
\end{equation}
where Faraday rotation has been neglected. 
The momentum differences $\Delta_{||,\perp}$ arise due to the effects of the propagation of photons in a plasma, QED vacuum polarisation, photon-photon dispersion in background radiation fields~\cite{2015PhRvD..91h3003D}, and possible photon absorption~\cite[e.g.,][]{2011PhRvD..84j5030D}, 
$\Delta_{\perp} = \Delta_\mathrm{pl} + 2\Delta_\mathrm{QED} + \chi - i\Gamma / 2$,  and $\Delta_{||} = \Delta_\mathrm{pl} + 7/2\Delta_\mathrm{QED} + \chi - i\Gamma / 2$.
The plasma term $\Delta_\mathrm{pl} = -\omega_\mathrm{pl} / (2 E)$ depends on the electron density $n_\mathrm{e}$ through 
the plasma frequency $\omega_\mathrm{pl}^2 = 4\pi n_e e^2 / m_e$, with the electric charge $e$ and electron mass $m_e$.
The QED vacuum polarisation is given by $\Delta_\mathrm{QED} = \alpha E / (45\pi)(B /(B_\mathrm{cr}))^2$, with the fine-structure constant $\alpha$, and the critical magnetic field $B_\mathrm{cr} = m^2_e / |e| \sim 4.4\times10^{13}\,$G.\footnote{In \gammaALPs, higher order terms of $B/B_\mathrm{cr}$ are incorporated as well using Eq. 6 in Ref.~\cite{2012ApJ...748..116P}.}
The photon dispersion term $\chi$ will depend on the energy density of the relevant background radiation fields. 
If only the cosmic microwave background (CMB) is of importance and the photons have energies far below the threshold of electron-positron pair production, it reads $\chi = 44\alpha^2 U_\mathrm{CMB} / (135 m_e^4) E$, where $U_\mathrm{CMB}$ is the total CMB energy density~\cite{2015PhRvD..91h3003D}. 
Photon absorption through, e.g., pair production, is given by the inverse mean free path (i.e., interaction rate) $\Gamma$.
The kinetic term for the ALP is $\Delta_a = -m_a^2 / (2E)$ and photon-ALP mixing is the result of the off-diagonal elements $\Delta_{a\gamma} = g_{a\gamma} B / 2$.
In \gammaALPs, the $\Delta$ terms are calculated in units of $\mathrm{kpc}^{-1}$ and the energy is assumed to be in GeV.
The ALP mass is stored in units of neV and the photon-ALP coupling in units of $10^{-11}\,\mathrm{GeV}^{-1}$.

For an unpolarised photon beam, Eq.~\eqref{eqn:eom} can be written in terms of the density matrix $\rho(x_3) = \Psi(x_3)\Psi(x_3)^\dagger$ that obeys the von-Neumann-like commutator equation $i\frac{\mathrm{d}\rho}{\mathrm{d}x_3} = [\rho,\mathcal{M}_0]$. 
It is solved through $\rho(x_3) = \mathcal{T}(x_3,0; E)\rho(0)\mathcal{T}^\dagger(x_3,0; E)$, with the transfer matrix 
$\mathcal T$ that solves Eq. \eqref{eqn:eom} with $\Psi(x_3) = \mathcal{T}(x_3,0;E)\Psi(0)$ and initial condition 
$\mathcal{T}(0,0;E) = 1$.
For the general case where $\psi\neq 0$,
the solutions have to be modified with a similarity transformation and, consequently, $\mathcal M$ and $\mathcal T$ will depend on $\psi$. 

Several models for different astrophysical magnetic field environments are implemented in \gammaALPs (see below) and the photon-ALP beam can traverse several of these environments from a source to the observer. 
For each environment $m$, we make the assumption that we can split the path into $N_m$ consecutive domains where in each domain the magnetic field, $\psi$, electron density, and dispersion terms $\chi$ are constant.
The transfer matrix of each environment $m=1,\ldots,M$, where $m=1$ denotes the environment closest to the source, is then found through matrix multiplication and the final transfer matrix from source to observer is given by
\begin{eqnarray}
\mathcal{T}_\mathrm{tot} &=& 
\mathcal{T}_{M}(x_{3,N_M},x_{3, N_{M-1}}; E)\cdots \mathcal{T}_{1}(x_{3,N_1},x_{3, 0}; E)\nonumber\\
&=& \prod\limits_{m=1}^M\prod\limits_{n = 1}^{N_m} \mathcal{T}_{M-m+1}(x_{3,N_{M-m+1} - n + 1},x_{3,N_{M-m+1} - n};E).
\end{eqnarray}
In the above equation, $x_{3,0}=0$, i.e. the coordinate closest to the source that produces a beam with initial polarization $\rho(0)$.
The total photon survival probability is then given by
\begin{equation}
P_{\gamma\gamma} = \mathrm{Tr}\left( (\rho_{11} + \rho_{22}) \mathcal{T}_\mathrm{tot} \rho(0)
\mathcal{T}_\mathrm{tot}^\dagger\right),
\label{eqn:surv-prob}
\end{equation}
with $\rho_{11} = \mathrm{diag}(1,0,0)$, $\rho_{22} = \mathrm{diag}(0,1,0)$, and $\rho(0) = \mathrm{diag}(1/2,1/2,0)$ for an initially unpolarized beam. 
Note that the transfer matrices adjacent to $\rho(0)$ in Eq.~\eqref{eqn:surv-prob} correspond to the domain closest to the source.

\section{Implemented astrophysical environments}

In Table~\ref{tab:environ}, we provide a list of the astrophysical environments currently implemented in \gammaALPs.
From source to the observer they include (i) two models at different levels of sophistication for AGN jets in which the $\gamma$-ray emission is produced;
(ii) two models for magnetic fields in galaxy clusters, one with a simple cell-like structure, the other one describes the $B$ field as a field with Gaussian turbulence;
(iii) a simple cell-like model for the intergalactic magnetic field (IGMF), which evolves with redshift and includes $\gamma$-ray absorption on the extragalactic background light (EBL);
(iv) the Galactic magnetic field of the Milky Way, where the user can choose between three different models provided in the literature, namely the ASS and BSS models in Ref.~\cite{2011ApJ...738..192P}, the model of Ref.~\cite{2012ApJ...757...14J}, or an updated version of the latter provided in Ref.~\cite{2016A&A...596A.103P}.
Additionally, an environment with pure EBL absorption and no photon-ALP mixing is provided, as well as the possibility to use externally computed $B$ fields and electron densities. 
Further details on the individual environments are provided in the respective references given in the table. 
More information can also be found in the \gammaALPs online documentation and tutorials.\footnote{\url{https://gammaalps.readthedocs.io/en/latest/tutorials/index.html}}

\begin{table}[htb]
    \begin{scriptsize}
    \centering
    \begin{tabular}{c|cccc}
    \hline
    \hline
    \textbf{Environment} & \multirow{2}{*}{\textbf{Description}}  & $B$\textbf{-field} & $n_\mathrm{el}$ & \multirow{2}{*}{\textbf{Ref.}} \\ 
    \textbf{Name} &  & \textbf{Model} & \textbf{Model} & \\
    \hline
    \hline
    \multirow{2}{*}{\texttt{Jet}} & Mixing in the toroidal magnetic field 
    & Power law in distance
    & Power law in distance
    & \multirow{2}{*}{\cite{2014JCAP...09..003M,2015PhLB..744..375T}} \\
     {} 
    &  of an AGN jet 
    & from central black hole
    & from central black hole
    & {} \\
    \hline
    \multirow{2}{*}{\texttt{JetHelicalTangled}} & Mixing in the helical and tangled 
    & Power law in distance
    & Power law in distance
    & \multirow{2}{*}{\cite{2021PhRvD.103b3008D}} \\
     {} 
    &  magnetic fields of an AGN jet 
    & from central black hole
    & from central black hole
    & {} \\
    \hline
    \multirow{4}{*}{\texttt{ICMCell}} & Mixing in a galaxy cluster magnetic field 
    & Constant in each cell
    & 
    & \multirow{4}{*}{\cite{2012PhRvD..86g5024H,2013PhRvD..87c5027M}} \\
     {} 
    & with a cell-like structure decreasing 
    &  with randomly varying $\psi$
    & (generalized) $\beta$ profile for
    & {} \\
     {} 
    & with growing distance from 
    & decreasing as
    &  galaxy clusters
    & {} \\
    {} 
    & cluster center following $n_\mathrm{el}(r)$
    & a function in $n_\mathrm{el}(r)$  
     & {} & {} \\
    \hline
    \multirow{4}{*}{\texttt{ICMGaussTurb}} & Mixing in a galaxy cluster magnetic field 
    & 
    & 
    & \multirow{4}{*}{\cite{2014JCAP...09..003M}} \\
     {} 
    &  with Gaussian turbulence decreasing 
    & Field with Gaussian
    & (generalized) $\beta$ profile for
    & {} \\
     {} 
    & with growing distance from 
    &   turbulence decreasing as
    &  galaxy clusters
    & {} \\
    {} 
    & cluster center following $n_\mathrm{el}(r)$
    &  a function in $n_\mathrm{el}(r)$  
    & {} & {} \\
    \hline
    \multirow{3}{*}{\texttt{IGMF}} & Mixing in the intergalactic magnetic field
    &  Constant in each cell 
    &  constant 
    & \multirow{3}{*}{\cite{2011PhRvD..84j5030D}} \\
    &  with a cell-like structure, 
    & with randomly varying $\psi$ 
    & (evolves with redshift)
    & {} \\
    & including EBL absorption
    & {}
    & {}
    & {} \\
    \hline
    \multirow{2}{*}{\texttt{GMF}} & Mixing in the coherent component of
    & \multirow{2}{*}{Models of Refs.~\cite{2011ApJ...738..192P,2012ApJ...757...14J,2016A&A...596A.103P}}
    & \multirow{2}{*}{constant}
    & \multirow{2}{*}{\cite{2012PhRvD..86g5024H}} \\
     {} 
    &  the magnetic field of the Milky Way 
    & 
    & 
    & {} \\
    \hline
    \multirow{2}{*}{\texttt{EBL}} & No mixing, only 
    & \multirow{2}{*}{--}
    & \multirow{2}{*}{--}
    & \multirow{2}{*}{--} \\
     {} 
    & photon absorption on EBL
    & 
    & 
    & {} \\
    \hline
    \multirow{2}{*}{\texttt{File}} & Environment initialized 
    & \multirow{2}{*}{As provided in the file}
    & \multirow{2}{*}{As provided in the file}
    & \multirow{2}{*}{--} \\
     {} 
    & from file
    & 
    & 
    & {} \\
    \hline
    \multirow{2}{*}{\texttt{Array}} & Environment initialized 
    & \multirow{2}{*}{As provided in the array}
    & \multirow{2}{*}{As provided in the array}
    & \multirow{2}{*}{--} \\
     {} 
    & from \texttt{numpy} array
    & 
    & 
    & {} \\
    \hline
    \end{tabular}
    \caption{Implemented environments in the \gammaALPs code. 
    References that describe the models for the magnetic fields and the electron densities are given in the last column. 
    }
    \label{tab:environ}
    \end{scriptsize}
\end{table}

\section{Workflow}

The \python code of a typical workflow example is shown below in Listing~\ref{lst:ex}, which demonstrates the basic features of the API of \gammaALPs.
This particular example assumes as a source NGC~1275, the central AGN of the Perseus galaxy cluster, and calculates mixing in the intra-cluster field and the Milky Way. No mixing in the IGMF is assumed and only EBL absorption is used. 
The same parameters as in Ref.~\cite{2016PhRvL.116p1101A} are chosen for the cluster field. 
After the basic imports from the \verb|gammaALPs.core| module, the user defines a $\gamma$-ray source with redshift and sky coordinates. The sky coordinates are necessary for $B$-field environments that depend on the sky position such as the Galactic magnetic field. 
Next, the user initializes the \verb|ALP| class, which stores the ALP mass and coupling to be used throughout the calculation. 
After setting the desired energy range and initial polarization matrix $\rho(0)$, the user then initializes the \verb|ModuleList| class which stores the different astrophysical environments. 
An environment is added with the \verb|add_propagation| function which takes the name of the environment (as listed in Table~\ref{tab:environ}), the positional index with respect to the source as inputs, and environment specific model parameters. The positional index 0 corresponds to the environment closest to the source. 
After the user has specified all desired environments, the conversion probability into the final polarization states is calculated by calling the \verb|run| function. 

\begin{lstlisting}[language=Python, caption=Python code example of the \gammaALPs workflow., label=lst:ex]
from gammaALPs.core import Source, ALP, ModuleList
import numpy as np

# define the source
ngc1275 = Source(z=0.017559, ra='03h19m48.1s', dec='+41d30m42s')

# define the ALP with mass in neV and coupling in 1e-11 GeV^-1
m, g = 1.,1.
alp = ALP(m,g)

# define an energy range in GeV
EGeV = np.logspace(1., 3.5, 250)

# define the initial polarization
pin = np.diag((1.,1.,0.)) * 0.5

# initialize the module list
ml = ModuleList(alp, ngc1275, pin=pin, EGeV=EGeV)
# add propagation modules
# first, mixing in Gaussian turbulent field
ml.add_propagation("ICMGaussTurb",
                  0, # position of module counted from the source.
                  nsim=10, # number of random B-field realizations
                  B0=10.,  # rms of B field in muG
                  # parameters for electron density, 
                  # see Churazov et al. 2003, Eq. 4
                  n0=3.9e-2,
                  n2=4.05e-3, 
                  r_core=80.,
                  r_core2=280.,
                  beta=1.2,
                  beta2=0.58,
                  r_abell=500., # extension of the cluster in kpc
                  eta=0.5, # scaling of B-field with electron density
                  kL=0.18, # maximum turbulence scale in kpc^-1
                  kH=9.,  # minimum turbulence scale
                  q=-2.80, # turbulence spectral index
                  seed=0 # random seed for reproducability
                 )
# EBL attenuation comes second, after beam has left cluster
ml.add_propagation("EBL",1, eblmodel='dominguez')
# finally, the beam enters the Milky Way Field
ml.add_propagation("GMF",2, model='jansson12')

# calculate the final polarization states
# for photon polarization along x and y
# and ALP polarization state
px, py, pa = ml.run()

# the total photon survival probability is 
# the sum of the final photon states
pgg = px + py
\end{lstlisting}

\begin{figure}
\centering
\includegraphics[width=0.32\linewidth]{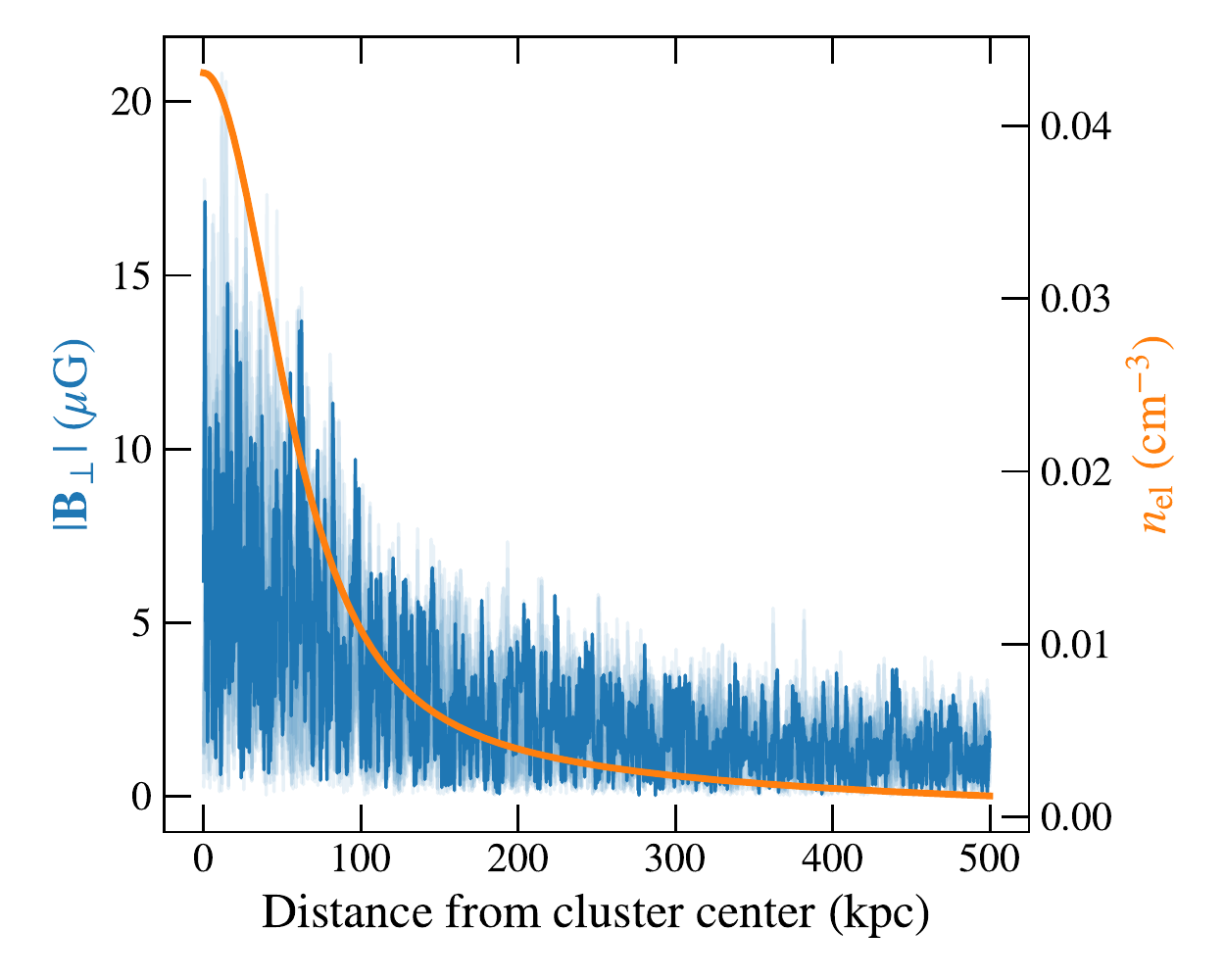}
\includegraphics[width=0.32\linewidth]{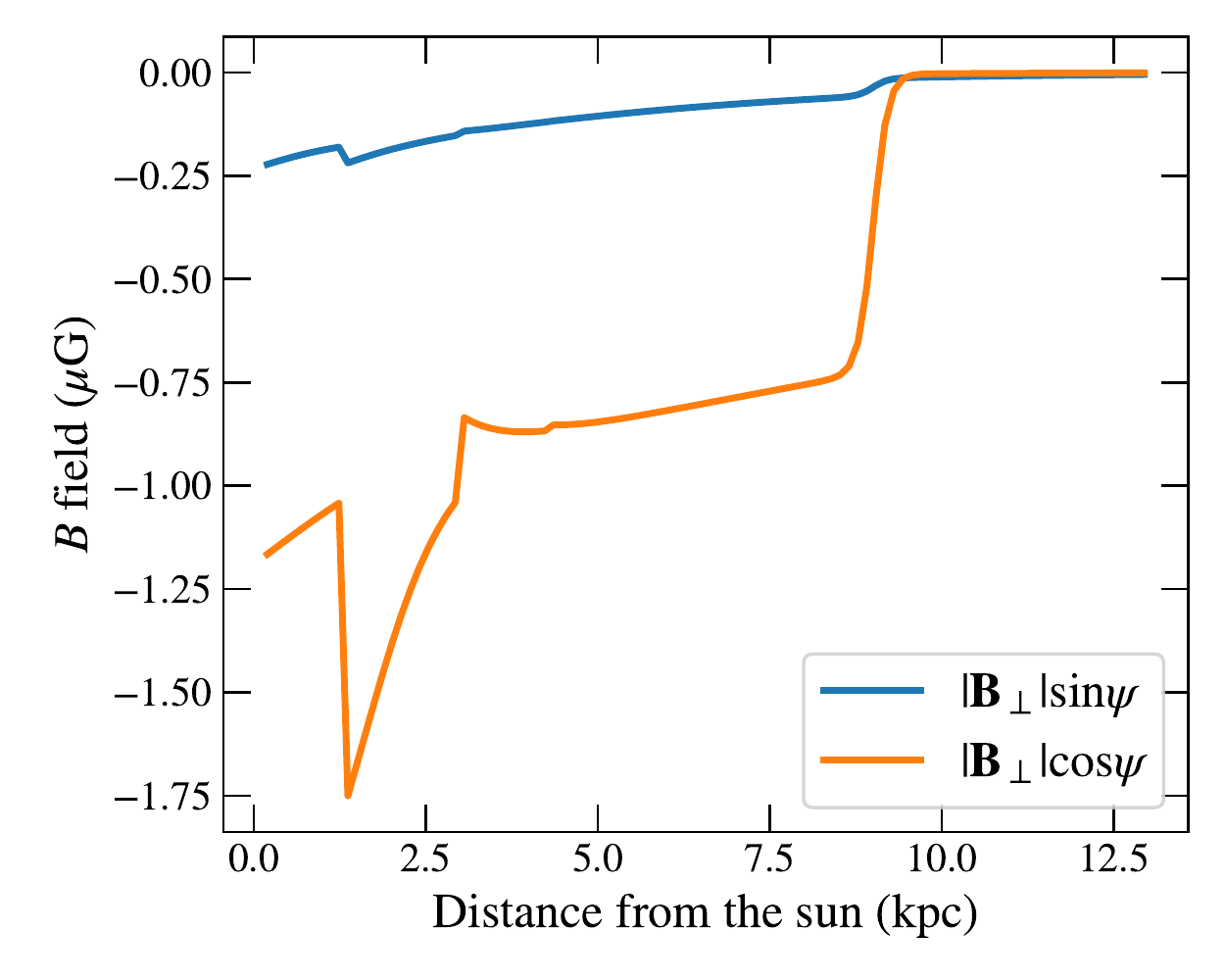}
\includegraphics[width=0.32\linewidth]{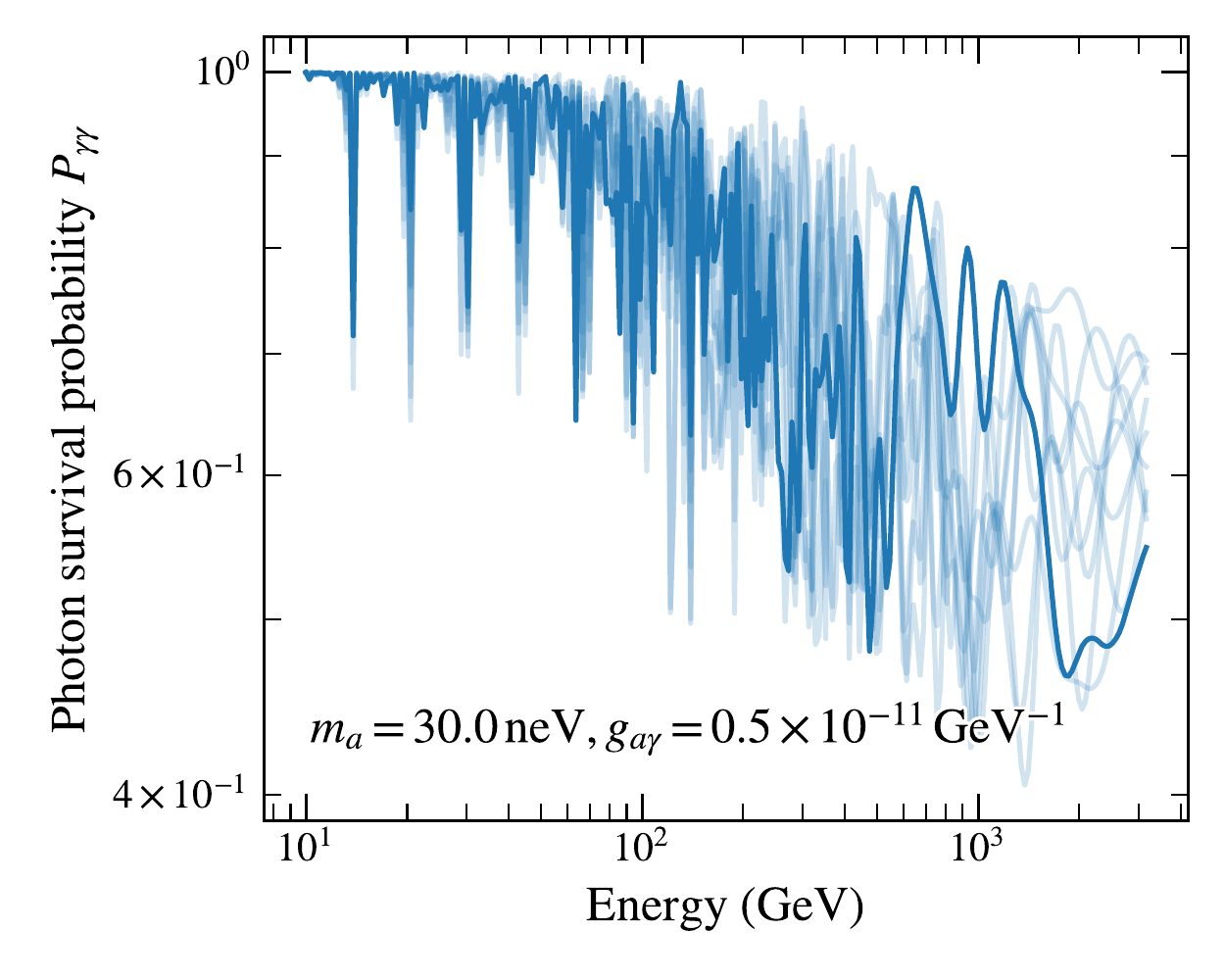}
\caption{Example of \gammaALPs environment models and final output.
\textit{Left:} 10 random realizations of the intra-cluster magnetic field as a function of the distance to the cluster center. The absolute value of the transversal magnetic field is shown with one realization highlighted as a solid blue line. The electron density is shown in orange (right $y$-axis). 
\textit{Center:} The transversal components of the Galactic magnetic field towards NGC~1275 is shown as a function of the distance from the sun. 
\textit{Right:} Resulting final photon survival probabilities for the 10 random $B$-field realizations. The same realization as in the left panel is highlighted with a solid blue line. 
}
\label{fig:mixing}
\end{figure}

Before calling \verb|run|, the user can also change the ALP parameters for which the mixing is calculated. For example, an ALP mass of 30\,neV and coupling of $g_{a\gamma}=5\times10^{-12}\,\mathrm{GeV}^{-1}$ is set by  \verb|ml.alp.m = 30.; ml.alp.g = 0.5|.
Furthermore, \verb|run| supports multiprocessing (using \python's built-in \verb|multiprocessing| package).
The actual calculation of the transfer matrices is sped up by using \numba~\cite{numba}.
We show the assumed magnetic field and the electron density as well as the final $P_{\gamma\gamma}$ in Figure~\ref{fig:mixing}. 
On a MacBook Pro with a 2.7\,GHz Intel Core i7 and 16GB RAM the calculation of $P_{\gamma\gamma}$ for this example (in total 4600 magnetic domains, 10 random realizations, 250 steps in energy) took $(17.7\pm 0.5)\,\mathrm{s}$. 

It is also worth mentioning that each environment can take a two-dimensional spline interpolation function as a keyword \verb|chi| for the photon-photon dispersion. For example, for the Galactic magnetic field, this could be an interpolated function for $\chi$ in terms of energy and distance traveled along the line of sight in the Milky Way, which describes the dispersion in the intestellar radiation fields~(this was used, e.g., in Ref.~\cite{2017arXiv171201839V}).
Further examples for other environments are provided in the \gammaALPs tutorials. 

\section{Conclusions}

We have presented the open-source \gammaALPs \python package, which computes photon-ALP oscillations in various astrophysical environments. 
The code features several models for astrophysical magnetic fields and electron distributions but it is also possible to provide custom $B$-field and electron density models.
As illustrated in these proceedings, with only a few lines of code it is possible to set up the models and run the calculation of the photon-ALP oscillation probability. 

Beyond the calculation of the photon-ALP conversion probability, the code can also be used for other astrophysical applications.
For example, the \python implementations of the Galactic field models can be useful for cosmic-ray propagation studies. 
It is also possible to calculate the rotation measure from the Gaussian turbulent magnetic field. 

In terms of future developments, it would be interesting to implement further models for astrophysical environments, such as a more realistic model for the IGMF, or dedicated models for the magnetic fields around magnetic white dwarfs and neutron stars. 
Also, further gains in terms of speed might be achieved through the efficient use of GPU programming. 

\begin{acknowledgments}
M.  M.  acknowledges  support from the European Research Council (ERC) under the European Union’s Horizon 2020 research and innovation program Grant agreement No. 948689 (AxionDM).
\end{acknowledgments}

\setstretch{1.}
\bibliographystyle{jhep}
\bibliography{references}

%
%
%

\end{document}